\documentclass[twocolumn,showpacs,preprintnumbers,pre,floatfix,superscriptaddress,amsmath,amssymb]{revtex4}
\usepackage{epsfig}
\usepackage{subfigure}
\usepackage{amsmath}
\usepackage{amssymb}
\usepackage{graphicx}
\usepackage{dcolumn}
\usepackage{bm}
\input epsf

\begin{document}

\title{The emergence of coherence in complex networks of heterogeneous dynamical systems}

\author{Juan G. Restrepo}
\email{juanga@math.umd.edu} \affiliation{ Institute for Research in
Electronics and Applied Physics, University of Maryland, College
Park, Maryland 20742, USA } \affiliation{ Department of Mathematics,
University of Maryland, College Park, Maryland 20742, USA }

\author{Edward Ott}
\affiliation{ Institute for Research in Electronics and Applied
Physics, University of Maryland, College Park, Maryland 20742, USA }
\affiliation{ Department of Physics and Department of Electrical and
Computer Engineering, University of Maryland, College Park, Maryland
20742, USA }

\author{Brian R. Hunt}
\affiliation{ Department of Mathematics, University of Maryland,
College Park, Maryland 20742, USA } \affiliation{ Institute for
Physical Science and Technology, University of Maryland, College
Park, Maryland 20742, USA}

\date{\today}

\begin{abstract}
We present a general theory for the onset of coherence in
collections of heterogeneous maps interacting via a complex
connection network. Our method allows the dynamics of the individual
uncoupled systems to be either chaotic or periodic, and applies
generally to networks for which the number of connections per node
is large. We find that the critical coupling strength at which a
transition to synchrony takes place depends separately on the
dynamics of the individual uncoupled systems and on the largest
eigenvalue of the adjacency matrix of the coupling network. Our
theory directly generalizes the Kuramoto model of equal strength,
all-to-all coupled phase oscillators to the case of oscillators with
more realistic dynamics coupled via a large heterogeneous network.

\end{abstract}


\maketitle

In recent years, much progress has been made in describing the
complex structure of real world networks \cite{newman1,barabasi1}.
The study of dynamical processes taking place in such complex
networks has applications in fields ranging from biology to
engineering. One of the most important phenomena involving networks
of coupled dynamical systems is the emergence of large-scale
coherent behavior \cite{pikovsky,mosekilde}. It is often observed
that large collections of heterogeneous dynamical systems (e.g.,
cells, fireflies) synchronize their rhythms so that a significant
proportion of the systems have states that are highly correlated
with those of the others. It is natural to ask what determines the
emergence of such coordinated behavior given the network of
interactions between the dynamical systems and their individual
dynamics.

The case of equal-strength, all-to-all coupled phase oscillators was
studied by Kuramoto \cite{kuramoto}, who considered the case of $N$
oscillators, each of which is described by a phase $\theta_j$ and an
intrinsic frequency $\omega_j$. Kuramoto assumed sinusoidal coupling
so that the phase of oscillator $j$ evolves as $ \dot{\theta}_j =
\omega_j + k \sum_{m=1}^N \sin(\theta_m -\theta_j). $ Kuramoto found
that, in the limit $N\to\infty$, for coupling strengths $k$ less
than a critical coupling strength $k_c$ that depends on the
distribution of frequencies, the phases of the
oscillators are incoherent, i.e., $\theta_j$ are uniformly
distributed on $[0,2\pi)$. For values of the coupling strength $k$
larger than $k_c$, a significant fraction of the oscillators evolve
with a common frequency. The Kuramoto model has become a classical
paradigm for the emergence of coherent behavior in an ensemble of
heterogeneous oscillators (see \cite{strogatz} for reviews).

Although the Kuramoto model has the advantage that it can be treated
analytically, it depends on the system being simple in two major
aspects. First, the network is assumed to be all-to-all, so that
every oscillator is coupled with uniform strength to every other
oscillator. Second, the dynamics and coupling term are highly
simplified: each oscillator is described only by its phase and the
coupling term is sinusoidal. Recently, collections of coupled
dynamical systems which have either a more general interaction
network or a more general dynamics have been studied. In previous
works \cite{onset,directed} (see also
\cite{ichinomiya,lee,ichinomiya2,jadbabaie,moreno}), we have studied
the Kuramoto phase oscillator model generalized to the case of a
general interaction network described by an adjacency matrix. We
found that there is still a transition to synchrony at a critical
coupling strength that depends on the largest eigenvalue of the
adjacency matrix and the distribution of frequencies.
On the other hand, `globally' coupled (i.e., equal
coupling strength, all-to-all) collections of many dynamical systems
with more general dynamics (e.g., mixed collections of chaotic and
periodic oscillators, chaotic maps, etc) have recently been studied
\cite{prk, sakaguchi, topaj, baek,ott}, and a transition to
synchrony has been observed at coupling strengths that depend on the
dynamics of the uncoupled systems.

Our aim in this Letter is to generalize these previous works by
studying large collections of heterogeneous general dynamical
systems coupled by networks with complex topology. We find that in
this general case there is a transition to coherence, and that the
coupling strengths at which it occurs can be obtained from the
uncoupled individual unit dynamics and the eigenvalues of the
adjacency matrix. Thus, we achieve a separation of the problem into
a part depending on the network only and a part depending on the
individual unit dynamics only \cite{pecora2}. Our model allows
strong heterogeneity and different dynamics in the collection of
dynamical systems, making it potentially appropriate to describe
biological or other strongly heterogeneous systems.

We study networks of $N$ coupled maps satisfying
\begin{equation}\label{eq1}
x_{n+1}^{(j)} = f(x_n^{(j)},\mu_j) + w_n^{(j)} +
\end{equation}
\begin{equation}
k g(x_n^{(j)})\sum_{m=1}^N A_{jm}[q(x_n^{(m)}) - \langle
q(x)\rangle]\nonumber.
\end{equation}
Here $j = 1, 2, \dots, N$ labels the map and we are interested in
large $N$, $f(x_n^{(j)},\mu_j)$ determines the uncoupled dynamics of
each individual map with a parameter $\mu_j$; for each map $j$ the
vector of parameters $\mu_j$ is chosen randomly and independently of
$j$ with a probability distribution $p(\mu)$; and the term
$w_n^{(j)}$ is a random noise which is assumed to be statistically
independent of $j$ and $n$ and to satisfy $E[w_n^{(j)}]= 0$,
$E[w_n^{(j)}w_m^{(l)}]= \sigma^2\delta_{nm}\delta_{jl}$, where
$E[\cdotp]$ represents the expected value. In the coupling term, $k$
is a global coupling strength and the scalar functions $g$ and $q$
are assumed to be smooth and bounded. The notation $\langle
\cdotp\rangle$ represents the average over the distribution of the
vector of parameters $\mu_i$ and over the natural measures of the
attractors of the noisy uncoupled ($k = 0$) system. Alternatively,
$\langle q \rangle$ is the $\mu$-average of the infinite time
average of $q(x_n)$ over a typical orbit $x_n$ of the noisy
uncoupled system. We remark that $\langle q \rangle$ is independent
of time. We also note that $x$ can be a vector for the situation of
multidimensional individual maps. For simplicity, in what follows we
consider $x$ to be a scalar.

The $N \times N$ matrix $A_{jm}$ determines the network of
interactions: node $m$ interacts with node $j$ only if $A_{jm} \neq
0$. We refer to those nodes $m$ for which $A_{jm} \neq 0$ as the
{\it neighbors} of node $j$, and to $d_j = \sum_{m=1}^N A_{jm}$ as
the {\it degree} of node $j$.

We are interested in studying system (\ref{eq1}) for the case in
which nodes have a large number of neighbors. In this case, if the
initial conditions are chosen distributed according to the natural
measure of the attractors of the uncoupled systems, then, because of
the large number of terms in the sum in the coupling term in
Eq.~(\ref{eq1}), the fact that $x_n$ are distributed according to
the measure of the uncoupled attractors, and the lack of
correlations between the parameter vectors and the network, we can
approximate
\begin{equation}\label{apro}
\addtolength{\abovedisplayskip}{-1mm}
\addtolength{\belowdisplayskip}{-1mm}
\sum_{m=1}^N A_{jm}q(x_n^{(m)}) \approx \langle q(x)\rangle
\sum_{m=1}^N A_{jm},
\end{equation}
and equality of these sums applies in the limit of an infinite
number of neighbors. We refer to this situation as the {\it
incoherent state}, and we will study in what follows its linear
stability. Under the previously mentioned assumption of a large
number of neighbors per node, the incoherent state is
(approximately) a solution of the system (\ref{eq1}). Its linear
stability can be studied using the same methods that were used in
Ref.~\cite{baek} for the all-to-all case. In the following, we will
adapt these techniques to the case of general connectivity.

In order to study the linear stability of the incoherent state, we
assume that $x_n^{(i)}$ is in the incoherent state and introduce an
infinitesimal perturbation $\delta x_n^{(i)}$. Linearization of
Eq.~(\ref{eq1}) produces
\begin{equation}\label{lin2}
\addtolength{\abovedisplayskip}{-2mm}
\addtolength{\belowdisplayskip}{-3mm}
\delta x_{n+1}^{(i)} = f'(x_n^{(i)},\mu_i) \delta x_{n}^{(i)} + k
g(x_{n}^{(i)})\sum_{j=1}^N A_{ij}q'(x_{n}^{(j)})\delta x_{n}^{(j)}.
\end{equation}
In order to solve Eq.~(\ref{lin2}) we consider (as in the variation
of parameters method for differential equations) a perturbation
$\epsilon_n^{(i)}$ of the uncoupled system $\epsilon_{n+1}^{(i)} =
f'(x_n^{(i)},\mu_i) \epsilon_{n}^{(i)}$ with $\epsilon_0 = 1$.
Defining $\Gamma_n^{(i)}=\sum_{j=1}^N A_{ij} q'(x_{n}^{(j)})\delta
x_{n}^{(i)}$ and assuming exponential growth, so that
$\Gamma_n^{(i)} = \gamma^{(i)}\eta^n$, we obtain for large $n$
\cite{footnote2}
\begin{equation}\label{before}
\addtolength{\abovedisplayskip}{-2mm}
\addtolength{\belowdisplayskip}{-3mm}
\gamma^{(m)} = k \sum_{i=1}^NA_{mi}\gamma^{(i)}\sum_{p=0}^n
\frac{q'(x_{n+1}^{(i)})\epsilon_{n+1}^{(i)}g(x_p^{(i)})\eta^{p-n-1}}{\epsilon_{p+1}^{(i)}}.
\end{equation}
In order to proceed further, we will again use the assumptions of
large number of neighbors per node and statistical independence of
the network and the vector of parameters. As we did in
Eq.~(\ref{apro}), we approximate Eq.~(\ref{before}) by
\begin{equation}\label{after}
\addtolength{\abovedisplayskip}{-2mm}
\addtolength{\belowdisplayskip}{-3mm}
\gamma^{(m)} = k
\left<\sum_{p=0}^n\frac{q'(x_{n+1}^{(i)})\epsilon_{n+1}^{(i)}g(x_p^{(i)})\eta^{p-n-1}}{\epsilon_{p+1}^{(i)}}\right>
\sum_{i=1}^NA_{mi}\gamma^{(i)}.
\end{equation}
[If the sum over $i$ in Eq.~(\ref{before}) is imagined as
approximating $N$ times the expected value of a product of two
random variables, Eq.~(\ref{after}) approximates $N$ times the
product of the two expected values as suggested by our assumption of
their independence.]

Let $Q(\eta)$ be the average in Eq.~(\ref{after}). With $m = n - p$,
and letting $n \to \infty$,
\begin{equation}\label{Qnoave}
Q(\eta) =
\eta^{-1}\left<\sum_{m=0}^{\infty}\frac{\epsilon_{n+1}}{\epsilon_{n-m+1}}q'(x_{n+1})g(x_{n-m})\eta^{-m}\right>.
\end{equation}
This quantity depends only on the dynamics of the individual
uncoupled oscillators, and also results from the analysis of the
globally coupled case \cite{baek}. From Eq.~(\ref{after}) we obtain
$ u^{(j)} = k_c \lambda_j Q(\eta)u^{(j)}$, where $u^{(j)}$ and
$\lambda_j$ denote, respectively, the eigenvectors of $A$ and their
corresponding eigenvalues. The onset of instability of the
incoherent state corresponds to $|\eta| = 1$, or $\eta =
e^{i\omega}$. Thus, network mode $j$ becomes unstable at a critical
coupling strength satisfying
\begin{equation}\label{result}
\addtolength{\abovedisplayskip}{-4mm}
\addtolength{\belowdisplayskip}{-4mm}
1 = k_c \lambda_j Q(e^{i\omega}),
\end{equation}
where the critical frequency $\omega$ is found by $Im[\lambda_j
Q(e^{i\omega})] = 0$, and $Im$ denotes the imaginary part. We are
interested in the solutions $k_c$ of Eq.~(\ref{result}) with the
smallest magnitude. Typically, they correspond to the mode
associated to the eigenvalue of largest magnitude, which is usually
real \cite{footnote}. In this case the critical frequency is found
from $Im[Q(e^{i\omega})] = 0$ and is independent of the network.

We remark that the definition and numerical computation of $Q(\eta)$
is in general nontrivial, since for $|\eta| = 1$ (necessary for
investigation of the onset of coherence) the individual terms in the
sum in Eq.~(\ref{Qnoave}) diverge with $m$ for typical initial
condition in chaotic maps. However, for large enough $|\eta|$ these
terms decay exponentially with increasing $m$ and consequently the
sum and average can be interchanged, so that
\begin{equation}\label{Qave}
\addtolength{\abovedisplayskip}{-1mm}
\addtolength{\belowdisplayskip}{-1mm}
Q(\eta) =
\eta^{-1}\sum_{m=0}^{\infty}\left<\frac{\epsilon_{n+1}}{\epsilon_{n-m+1}}q'(x_{n+1})g(x_{n-m})\right>\eta^{-m}.
\end{equation}
In Refs.~\cite{baek,ott} it is argued that the averages in the
summand of (\ref{Qave}) decay exponentially with $m$, so that
(\ref{Qave}) can be analytically continued to $|\eta| = 1$ as
desired.

We note that the function $Q(\eta)$ depends exclusively on the
dynamics of the uncoupled oscillators and their parameter
distribution function $p(\mu)$, while the eigenvalues $\lambda_j$
depend exclusively on the network. Therefore, an independent
treatment of these problems allows determination of the critical
coupling strengths for the full system given by Eq.~(\ref{eq1}).

We now present two examples illustrating our theory. In order to
quantify the coherence, we define an order parameter $r$ by
\begin{equation}\label{r}
\addtolength{\abovedisplayskip}{-2mm}
\addtolength{\belowdisplayskip}{-1mm}
r^2 = \left< \left(\frac{\sum_{m=1}^N \overline{d}_{m}[q(x_n^{(m)})
-
 \langle q(x)\rangle]}{\sum_{m=1}^N \overline{d}_{m}}\right)^2\right>_t,
\end{equation}
where $<\cdotp>_t$ denotes a time average and the in-degree
$\overline{d}_j$ is defined by $\overline{d}_m = \sum_{j=1}^N
A_{jm}$. Note that the numerator can be written as $ \sum_{j=1}^N
\sum_{m=1}^N A_{jm}[q(x_n^{(m)}) - \langle q(x)\rangle], $ and,
therefore, $r$ measures the $rms$ of the coupling term in
Eq.~(\ref{eq1}) [aside from the factor $g(x^{(j)})$]. Thus, the
incoherent state corresponds to $r\approx 0$. We will investigate
what happens to $r$ as the coupling strength $k$ is increased past
the critical values predicted by the theory.

In our numerical experiments, we compute the order parameter $r$
using Eq.~(\ref{r}) with $x_n^{(m)}$ obtained from iteration of
Eq.~(\ref{eq1}). We calculate the time average using $1000$
iterations after the initial transients have disappeared.
\begin{figure}[h]
\begin{center}
\epsfig{file = 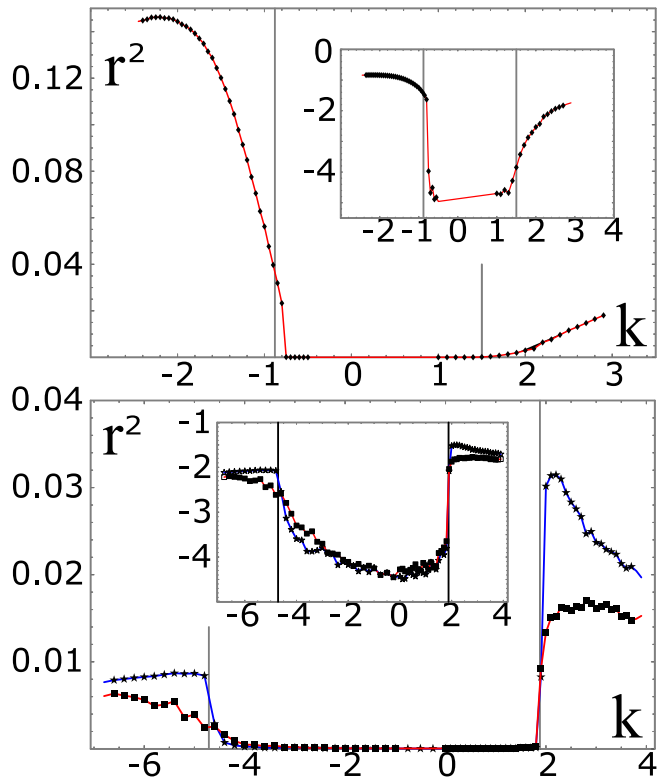, clip =  ,width=1.0\linewidth}
\addtolength{\abovecaptionskip}{-4mm}
\addtolength{\belowcaptionskip}{-8mm}
\caption{
The order parameter, $r^2$, plotted on a linear scale as a function
of the coupling strength $k$. The insets show the same data plotted
on a logarithmic scale for $r^2$. (a) Example 1 (identical noisy
maps) with a scale-free network with $N = 10^5$, exponent $-3$ and
$d_{min} = 100$. (b) Example 2 (heterogeneous noiseless maps) with a
scale-free network with $N = 20000$, exponent $-2.5$, and and lower
cutoff $d_{min} = 50$ (boxes) and $d_{min} = 200$ (stars). The
vertical lines indicate the theoretical values for the critical
coupling strength. } \label{figi23}
\end{center}
\end{figure}
As an example, we consider \cite{baek} for the functions in
Eq.~(\ref{eq1}),
$f(x_n^{(i)},\mu_i) = 2x_n^{(i)} + \mu_i$,
$q(x) = \cos x$, and $g(x) = \sin(2x) + \sin(4x)$.
Throughout  $x$ is regarded as an angle-like variable and its value
modulo $2\pi$ to be taken where appropriate. In Ref.~\cite{baek}
$Q(e^{i\omega})$ was calculated for Gaussian noise with mean zero
and standard deviation $\sigma$ to be
\begin{equation}\label{qbaek}
Q(e^{i\omega})= -\frac{1}{2}\left(
\frac{e^{-\sigma^2/2}}{e^{i\omega}}\langle\cos(\mu)\rangle + 2
\frac{e^{-5\sigma^2/2}}{e^{2i\omega}}\langle\cos(3 \mu)\rangle
\right),
\end{equation}

We numerically consider the following two examples, listed below
with the theoretical critical coupling strengths obtained using
expression (\ref{qbaek}) in Eq.~(\ref{result}).
\begin{enumerate}
\item Identical noisy maps with $\sigma = 0.4$, $p(\mu) = \delta(\mu)$.
For this example the solutions of Eq.~(\ref{result}) are $k_c
\lambda_j \approx 1.49$ and $k_c\lambda_j \approx -0.88$.

\item Heterogeneous noiseless maps with $\sigma = 0$,
$p(\mu) = 2/\pi$ if $0\leq \mu < \pi/2$, $0$ otherwise.  In this
example we obtain $k_c\lambda_j = 3\pi/5$ and $k_c\lambda_j =
-3\pi/2$.
\end{enumerate}
For the network connectivity, we consider a scale free network with
exponent $\gamma$, i.e., a network in which the degree distribution
$P(d)$ satisfies $P(d)\propto d^{-\gamma}$ for $d \geq d_{min}$ and
$0$ otherwise. We impose a lower cutoff $d \geq d_{min}$ so that our
assumption of a large number of neighbors per node is satisfied. We
use $\gamma = 3$, $d_{min} = 100$ for example $1$ and $\gamma =
2.5$, $d_{min} = 50$ and $d_{min} = 200$ for example $2$. In order
to construct such networks we use the Random Graph model of Chung
{\it et al.} \cite{chung}.

For scale free networks as described above, the largest positive
eigenvalue $\lambda_+$ is significantly larger than the magnitude of
the most negative eigenvalue $\lambda_-$, and so the mode
desynchronizing first is the one associated to the largest positive
eigenvalue. Consequently, the value of $\lambda_j$ used to determine
the critical coupling strengths $k_{c}$ for these examples will be
$\lambda_+$.

In Figs.~\ref{figi23}(a) and (b) we show the square of the order
parameter as a function of $k$ for examples $1$ and $2$ with $N =
10^5$ and $N = 20000$, respectively (the insets show the same
quantity on a logarithmic scale). For example $1$
[Fig.~\ref{figi23}(a)] the order parameter grows for values of the
coupling strength close to the positive and negative critical values
predicted by the theory (vertical lines in the figures). For example
$2$ [Fig.~\ref{figi23}(b)] the transition on the positive side is
quite sharp and occurs very close to the theoretical value $k_{c1}
\approx 1.88$, while on the negative side the transition, although
somewhat less well defined, also occurs close to the theoretical
critical coupling strength $k_{c2}\approx -4.71$. On the negative
side, the transition is not so sharp. However, we observe that as we
increase $d_{min}$ from $50$ (boxes) to $200$ (stars), the
transition becomes sharper. Generally, we find that the agreement
with the theoretical results improves as the minimum degree
$d_{min}$ and $N$ become larger \cite{footnote3}. The plots in
Fig.~\ref{figi23} were produced by starting in the incoherent state
and increasing the magnitude of the coupling strength $k$ from zero.
We have found that the transitions in our examples are not
hysteretic, i.e., the same behavior is observed if we use the
synchronized state as an initial condition and decrease the
magnitude of the coupling strength. We note that in the globally
coupled case, it was found that the computation of $Q(\eta)$ fails
to converge for an ensemble of identical noiseless logistic maps
\cite{baek}. It was argued that this results from structural
instability of the map and singularities in its invariant density,
which make a perturbation approach questionable. Since the
definition and numerical determination of $Q(\eta)$ in our case and
for the globally coupled case are identical, the lack of convergence
observed for this example in the globally coupled case will also
occur in the case of a network. However, we note that a small amount
of either noise or parameter heterogeneity was shown in \cite{baek}
to restore the validity of the results.

In summary, we have studied the onset of synchronization in large
networks of coupled maps (the case of coupled continuous time
oscillators can also be treated by these methods and will be
discussed elsewhere). We have found that the critical coupling
strength at which the transition to synchrony takes place depends
separately on the dynamics of the individual uncoupled oscillators
and on the largest eigenvalue of the adjacency matrix of the
network. Thus, we have achieved a separation of the problem of the
stability of the incoherent state for networks of coupled dynamical
systems into a part depending on the dynamics of the uncoupled
individual units and a part depending exclusively on the network
\cite{pecora2}. Our theory directly generalizes the Kuramoto model
of equal strength, all-to-all coupled phase oscillators to the case
of oscillators with more realistic dynamics coupled in a potentially
complex network. The results we obtain suggest that knowledge of
network properties that favor larger maximum eigenvalues can be used
to promote synchronism in large networks of coupled dynamical
systems.

This work was sponsored by ONR (Physics) and by NSF (DMS 0104087 and
PHY 0456240).


\begin{thebibliography}{99}
\bibitem{footnote2} The derivation follows closely the all-to-all case in Ref.~\cite{baek},
which assumes $A_{nm} = 1$. Further details will be published
elsewhere.
\bibitem{footnote} For symmetric matrices or matrices with positive entries
the eigenvalue with largest magnitude is real. For a large class of
nonsymmetric matrices with mixed positive/negative entries, we have
found that if the average of the nonzero elements of the matrix is
large enough, the eigenvalue with largest magnitude is also real and
it is well separated from the eigenvalue with next largest
magnitude. For details, see the Appendix of Ref.~\cite{directed}.
\bibitem{footnote3} A quantitative treatment and theory for the effect
of $d_{min}$ is given in \cite{onset} for the case of the Kuramoto model,
and we expect similar behavior when the dynamics of $f$ at different parameter
values are sufficiently similar.
\end{thebibliography}
\end{document}